\documentclass[aps,preprint,superscriptaddress,showpacs,nofootinbib]{revtex4-1}
\pdfoutput=1 

\usepackage[T1]{fontenc} 
\usepackage[compat=1.1.0]{tikz-feynman}
\usepackage{pifont}
\usepackage[mathscr]{eucal}
\usepackage{multirow}

\usepackage{graphicx}
\usepackage{amsfonts}
\usepackage{amsmath}
\usepackage{amssymb}
\usepackage{epsfig}
\usepackage{slashed}
\usepackage{dcolumn}%

\usepackage{ulem}
\usepackage{color}
\usepackage{caption}
\usepackage{subcaption}
\usepackage{setspace}
\usepackage[colorlinks,
linkcolor=blue,
anchorcolor=blue,
citecolor=green
]{hyperref}
\begin{document}
\title{The $Z$ resonance, inelastic dark matter, and new physics anomalies in the 
Simple Extension of the Standard Model (SESM) with general scalar potential}


\author{Wenxing Zhang}
\email{zhangwenxing@sjtu.edu.cn}
\affiliation{Tsung-Dao Lee Institute and School of Physics and Astronomy,
Shanghai Jiao Tong University, 800 Dongchuan Rd., Minhang, Shanghai 200240, China}

\author{Tianjun Li}
\email{tli@itp.ac.cn}
\affiliation{CAS Key Laboratory of Theoretical Physics, Institute of Theoretical Physics,
Chinese Academy of Sciences, Beijing 100190, China}
\affiliation{School of Physical Sciences, University of Chinese Academy of Sciences,
No.~19A Yuquan Road, Beijing 100049, China}

\author{Xiangwei Yin}
\email{yinxiangwei@itp.ac.cn}
\affiliation{CAS Key Laboratory of Theoretical Physics, Institute of Theoretical Physics,
Chinese Academy of Sciences, Beijing 100190, China}
\affiliation{School of Physical Sciences, University of Chinese Academy of Sciences,
No.~19A Yuquan Road, Beijing 100049, China}


\begin{abstract}
\begin{spacing}{1.33}
We consider the generic scalar potential with CP-violation, and study the $Z$ resonance
and inelastic dark matter in the Simple Extension of 
the Standard Model (SESM), which can explain the dark matter as well as
 new physics anomalies such as the B physics anomalies and muon anomalous magnetic moment, etc.
With the new scalar potential terms, we obtain the mass splittings for the real and imaginary parts of scalar fields.
And thus we can have the DM co-annihilation process mediated by $Z$ boson, which couples exclusively  
to the CP-even and CP-odd parts of scalar fields. This is a brand new feature compared to the previous study.
 For the CP conserving case, we present the viable parameter space for the Higgs and $Z$ resonances, 
which can explain the B physics anomalies, muon anomalous magnetic moment, and dark matter relic density, as well as
evade the constraint from the XENON1T direct detection simultaneously. 
 For the CP-violating case, we consider the inelastic dark matter, and study four concrete scenarios for 
the inelastic DM-nucleon scatterings mediated by the Higgs and $Z$ bosons in details. 
Also, we  present the benchmark points which satisfy the aforementioned constraints.
Furthermore, we investigate the constraints from the dark matter-electron inelastic scattering processes 
 mediated by the Higgs and $Z$ bosons in light of the XENONnT data. We show that
the constraint on the $Z$ mediated process is weak, while the Higgs mediated process excludes 
the dark matter with mass around several MeV.
\end{spacing}
\end{abstract}

\maketitle
\flushbottom
\section{Introduction}\label{Section1}

The celebrated theory known as the Standard Model (SM) of particle physics has been confirmed to be an effective description of our nature at the low energy scale after the  discovery of Higgs boson at the LHC in 2012~\cite{ATLAS:2012yve,CMS:2012qbp}. However, some inconsistence phenomena are much discomforting, for instance,  dark matter (DM), dark energy,  neutrino masses and mixings, 
and matter-antimatter asymmetry, etc. Besides, there exist a few big fine-tuning problems, for example, 
cosmological constant problem, gauge hierarchy problem, and strong CP problem, etc. 
Thus, we need to explore the new physics beyond the SM.

One of the pressing extensions is in flavour sector. The LHCb Collaboration has observed the persistent discrepancies
 between the SM predictions and experimental measurements for rare decays of B mesons in several years,
for example,  the angular distribution of $B\rightarrow K \mu^{+} \mu^{-}$ and Lepton Flavor Universality (LFU) ratios $R_{K^{(*)}}=BR(B \rightarrow K^{(*)}\mu \mu)/BR(B \rightarrow K^{(*)}e e)$~\cite{LHCb:2021trn,Altmannshofer:2021qrr,Aebischer:2019mlg,Albrecht:2018vsa,Li:2018lxi,Bifani:2018zmi,Arnan:2016cpy,Arnan:2019uhr,Arnan:2016cpy,Arnan:2019uhr,Yin:2020afe}. 
The recent measurements of $B^{+} \rightarrow K^{+} \ell^+ \ell^-$ and 
$B^{0} \rightarrow K^{*0} \ell^+ \ell^-$ decays have been presented to  
test the muon-electron universality in two ranges of the square 
of the dilepton invariant mass~\cite{LHCb:2022qnv,LHCb:2022zom}.
The results seems compatible with the SM predicitions.
However, this is not the final result, since this is just one measurement with the limited precision. In fact, the data used in such analyses are only few percents of LHC data. In addition, although the misidentification backgrounds for electron channel were underestimated in the previous measurement of 
$R_K/R_{K^*}$, the measurements for the muon channels are still correct. 
Also, for the $b\rightarrow s \mu^+ \mu^-$ differential decay branching ratio, there still exist
the theoretical and experimental deviations in the low $q^{2}$ region, and $P^{\prime}_{5}$, etc. Thus, the B physics anomalies are still worth considering.
The measurements of the anomalous magnetic moment of muon, $a_{\mu}=\left(g-2\right)_{\mu}/2$, is one of the most important directions to probe new physics (NP). The state-of-the-art measurements declared by the Fermilab experiment 
shows 4.2 $\sigma$ discrepancy between the SM prediction and experimental value~\cite{Muong-2:2021ojo}. This makes muon g-2 an intriguing topic in the future~\cite{Lindner:2016bgg, Capdevilla:2020qel,Buttazzo:2020ibd,Capdevilla:2021rwo,Li:2021poy,Ahmed:2021htr,Zhu:2021vlz,Calibbi:2021qto,Arcadi:2021cwg}. Moreover, there exist compelling evidences 
for the DM existence from both particle physics and astronomy. 
However, there exists a wide range of mass for the DM candidates, making DM physics a fruitful theme.
for a review, see~\cite{Bertone:2004pz}. Futhermore,
The XENON1T experiment has found a low-energy electron recoil signal about $1\sim7$ keV~\cite{XENON:2020rca}. But the signal was excluded by the XENONnT experiment~\cite{Aprile:2022vux}, setting new constraints on
various models.

In this paper, we consider the general scalar potential with CP-violation and the inelastic dark matter
 in the Simple Extension of the Standard Model (SESM)~\cite{Calibbi:2019bay,Li:2022eby}.
In this model, we can address $R_{K}$, $B_{s}-\bar{B}_{s}$ oscillation, muon anomalous magnetic moment, 
dark matter, and evade the XENON1T direct detection simultaneously, etc.
With the new scalar potential terms, we obtain the mass splittings for the real and imaginary parts of scalar fields.
And thus we can have the DM co-annihilation process mediated by $Z$ boson, which couples exclusively  
to the CP-even and CP-odd parts of scalar fields. This is a brand new feature compared to the Ref.~\cite{Li:2022eby}.
 For the CP conserving case, we present the viable parameter space for the Higgs and $Z$ resonances, 
which can explain the B physics anomalies, muon g-2, and DM relic density, as well as
evade the constraint from the XENON1T direct detection simultaneously. 
 For the CP-violating case, we discuss the inelastic dark matter, and consider four concrete scenarios for 
the inelastic DM-nucleon scatterings mediated by the Higgs and $Z$ bosons in details. Also, we 
 present the benchmark points which satisfy the aforementioned constraints.
Furthermore, we investigate the constraints from the dark matter-electron inelastic scattering processes 
 mediated by the Higgs and $Z$ bosons in light of the XENONnT data. We show that
the constraint on the $Z$ mediated process is weak, while the Higgs mediated process excludes 
the dark matter with mass around several MeV.

This paper is organized as follows. In Section~\ref{Section2}, we introduce the SESM,
 and discuss the scalar masses. In Section~\ref{Section3}, we address the new physics anomalies 
and DM as mentioned above. In Section~\ref{Section4}, four types of inelastic DM scenarios are discussed. 
In Section~\ref{Section5}, the constraints to our model by considering the null excess signals of XENONnT 
data are studied. We conclude in Section~\ref{Section6}.

\section{The Simple Extension of the Standard Model (SESM) }\label{Section2}

The SESM~\cite{Calibbi:2019bay,Li:2022eby} has been proposed to address the tentative 
new physics anomalies and DM in its own right. The model introduces a singlet complex scalar $\Phi_{S}$, a  doublet complex scalar $\Phi_{D}$, a vectorlike quark $Q^\prime$, and a vectorlike lepton $L^\prime$. All these exotic fields are odd under a discrete $Z_{2}$ symmetry while the SM field are even. The quantum numbers of additional fields are
\begin{equation}
\begin{array}{ccccc}
\hline \text { Field } & \text { spin } & S U(3)_{C} & S U(2)_{L} & U(1)_{Y} \\
\hline Q^{\prime} & 1 / 2 & \mathbf{3} & \mathbf{2} & 1 / 6 \\
L^{\prime} & 1 / 2 & \mathbf{1} & \mathbf{2} & -1 / 2 \\
\Phi_{S} & 0 & \mathbf{1} & \mathbf{1} & 0 \\
\Phi_{D} & 0 & \mathbf{1} & \mathbf{2} & -1 / 2 \\
\hline
\end{array}~.~
\end{equation}

The new fields can be written as
\begin{equation}
Q^{\prime}=\left(\begin{array}{c}
U^{\prime} \\
D^{\prime}
\end{array}\right)~, \quad L^{\prime}=\left(\begin{array}{c}
L^{\prime 0} \\
L^{\prime-}
\end{array}\right)~, \quad \Phi_{S} \equiv S_{s}^{0}~, \quad \Phi_{D}=\left(\begin{array}{c}
S_{d}^{0} \\
S^{-}
\end{array}\right)~.
\end{equation}

The Lagrangian is given by

\begin{align}
\mathcal{L} \supset &\left(\lambda_{i}^{Q} \overline{Q^{\prime}} Q_{i} \Phi_{S}+\lambda_{i}^{U} \overline{Q^{\prime}} U_{i} \Phi_{D}+\lambda_{i}^{D} \overline{Q^{\prime}} D_{i} \widetilde{\Phi}_{D}+\lambda_{i}^{L} \overline{L^{\prime}} L_{i} \Phi_{S}+\lambda_{i}^{E} \overline{L^{\prime}} E_{i} \widetilde{\Phi}_{D}+a_{H} H^{\dagger} \widetilde{\Phi}_{D} \Phi_{S}\right. \nonumber\\
&\left.+a_{H}^{\prime} H^{\dagger} \widetilde{\Phi}_{D} \Phi_{S}^{\dagger} + \lambda_{DH}^{\prime\prime} \left({\widetilde{\Phi}_{D}}^{\dagger} H\right)^{2}+ \lambda_{SH}^{\prime} \Phi_{S}^{2}|H|^{2} + \text { h.c. }\right) -M_{Q} \overline{Q^{\prime}} Q^{\prime}-M_{L} \overline{L^{\prime}} L^{\prime}-M_{S}^{2} |\Phi_{S}|^{2}\nonumber\\
&+ {M_{S}^{\prime}}^{2}\left(\Phi_{S}^{2}+{\Phi_{S}^{\star}}^{2}\right)-M_{D}^{2}|\Phi_{D}|^{2}-M_{H}^{2}|H|^{2}+ \frac{\lambda_{S}}{2}\left(\Phi_{S}^{+} \Phi_{S}\right)^{2}+ \lambda_{S}^{\prime}\left(\Phi_{S}^{2}+{\Phi_{S}^{\star}}^{2}\right)|\Phi_{S}|^{2} \nonumber\\
&+\lambda_{S}^{\prime\prime}\left(\Phi_{S}^{4}+{\Phi_{S}^{\star}}^{4}\right)+\frac{\lambda_{D}}{2}\left(\Phi_{D}^{+} \Phi_{D}\right)^{2} + \frac{\lambda_{H}}{2}\left|H\right|^{4}+ \lambda_{SD}\left|\Phi_{S}\right|^{2}\left|\Phi_{D}\right|^{2}+\lambda_{SD}^{\prime}\left(\Phi_{S}^{2}+{\Phi_{S}^{\star}}^{2}\right)|\Phi_{D}|^{2}\nonumber\\
&+\lambda_{S H}\left|\Phi_{S}\right|^{2}|H|^{2}+\lambda_{DH}\left|\Phi_{D}\right|^{2}|H|^{2}+ \lambda_{DH}^{\prime} \left(H^{\dagger} \widetilde{\Phi}_{D}\right)\left(\widetilde{\Phi}_{D}^{\dagger} H\right)~, \label{lagrangian}
\end{align}
where $H$ is the SM Higgs field, and
we denote the left-handed quark doublets, 
right-handed up-type quarks, right-handed down-type quarks, left-handed lepton doublets, and right-handed down-type leptons as  $Q_{i}$, $U_{i}$, $D_{i}$, $L_{i}$, and $E_{i}$ ($i$=1,2,3), respectively. 

After the electroweak symmetry breaking (EWSB), the above $\lambda_{SH}$ and $\lambda_{SH}^{\prime}$ terms
 will contribute to the mass of $\Phi_{S}$, while the $\lambda_{DH}$, $\lambda_{DH}^{\prime}$, and $\lambda_{DH}^{\prime \prime}$ terms will contribute to the mass of $\Phi_{D}$. Especially, we have the real and imaginary parts splittings 
of new scalars due to the  $a_{H}$, $a_{H}^{\prime}$ ${M_{S}^{\prime}}^{2}$, $\lambda_{SH}^{\prime}$ and $\lambda_{DH}^{\prime\prime}$ terms. In the next Section, we can obtain the inelastic DM co-annihilations through the $Z$ resonance 
because of such splittings.

Assume that CP is conserved, we obtain
the  mass square matrix of the real parts of $\Phi_{S}$ and $\Phi_{D}$ 
\begin{equation}
    \left(\begin{array}{cc}
{M_{S}}^{2}-2 {M_{S}^{\prime}}^{2}+\frac{v^{2} \lambda_{SH}}{2}-v^{2} \lambda_{SH}^{\prime} & \frac{(a_{H}^{\prime}-a_{H}) v}{\sqrt{2}} \\
\frac{(a_{H}^{\prime}-a_{H}) v}{\sqrt{2}} & M_{D}^{2}+\frac{v^{2} \lambda_{DH}}{2}-\frac{v^{2} \lambda_{DH}^{\prime}}{2}-v^{2} \lambda_{DH}^{\prime \prime}
\end{array}\right)~,
\label{Real}
\end{equation}
and the mass square matrix of their imaginary parts is
\begin{equation}
    \left(\begin{array}{cc}
{M_{S}}^{2}+2 {M_{S}^{\prime}}^{2}+\frac{v^{2} \lambda_{SH}}{2}+v^{2} \lambda_{SH}^{\prime} & \frac{(a_{H}+a_{H}^{\prime}) v}{\sqrt{2}} \\
\frac{(a_{H}+a_{H}^{\prime}) v}{\sqrt{2}} & M_{D}^{2}+\frac{v^{2} \lambda_{DH}}{2}-\frac{v^{2} \lambda_{DH}^{\prime}}{2}+v^{2} \lambda_{DH}^{\prime \prime}
\end{array}\right),
\label{Imaginary}
\end{equation}
where $v\simeq$ 246 GeV is the vacuum expectation value of the Higgs field.
And the corresponding mass eigenvalues of real parts are
\begin{equation}
    \begin{split}
       {M_{S_{1}}}^{2}&=\frac{1}{4}\left(2 M_{D}^{2}+2 M_{S}^{2}-4 {M_{S}^{\prime}}^{2}+v^{2}\left(\lambda_{DH}-\lambda_{DH}^{\prime}-2 \lambda_{DH}^{\prime \prime}+\lambda_{SH}-2 \lambda_{SH}^{\prime}\right)-\sqrt{A
       +B}\right)~,\\
   {M_{S_{2}}}^{2}&=\frac{1}{4}\left(2 M_{D}^{2}+2 M_{S}^{2}-4 {M_{S}^{\prime}}^{2}+v^{2}\left(\lambda_{DH}-\lambda_{DH}^{\prime}-2 \lambda_{DH}^{\prime \prime}+\lambda_{SH}-2 \lambda_{SH}^{\prime}\right)+\sqrt{A
       +B}\right)~,\\
    \end{split}
\end{equation}
where A and B are defined as
\begin{equation}
    \begin{split}
A&=4\left(M_{D}^{2}-M_{S}^{2}+2{M_{S}^{\prime}}^{2}\right)^{2}+8(a_{H}-a_{H}^{\prime})^{2} v^{2}~,\nonumber\\
B&=v^{2}\left(\lambda_{DH}-\lambda_{DH}^{\prime}-2 \lambda_{DH}^{\prime \prime}-\lambda_{SH}+2 \lambda_{SH}^{\prime}\right)\left(4\left(M_{D}^{2}-M_{S}^{2}+2 {M_{S}^{\prime}}^{2}\right)\right.\nonumber\\
&\left.+v^{2}\left(\lambda_{DH}-\lambda_{DH}^{\prime}-2 \lambda_{DH}^{\prime \prime}-\lambda_{SH}+2 \lambda_{SH}^{\prime}\right)\right)~.
    \end{split}
\end{equation}
The corresponding mass eigenvalues of imaginary parts are
\begin{equation}
    \begin{split}
       M_{S_{1}^{\prime}}^{2}&=\frac{1}{4}\left(2 M_{D}^{2}+2 M_{S}^{2}+4 {M_{S}^{\prime}}^{2}+v^{2}\left(\lambda_{DH}-\lambda_{DH}^{\prime}+2 \lambda_{DH}^{\prime \prime}+\lambda_{SH}+2 \lambda_{SH}^{\prime}\right)-\sqrt{C
       +D}\right)~,\\
        M_{S_{2}^{\prime}}^{2}&=\frac{1}{4}\left(2 M_{D}^{2}+2 M_{S}^{2}+4 {M_{S}^{\prime}}^{2}+v^{2}\left(\lambda_{DH}-\lambda_{DH}^{\prime}+2 \lambda_{DH}^{\prime \prime}+\lambda_{SH}+2 \lambda_{SH}^{\prime}\right)+\sqrt{C
       +D}\right)~,\\
    \end{split}
\end{equation}
where C and D are defined as
\begin{equation}
    \begin{split}
C&=4\left(M_{S}^{2}-M_{D}^{2}+2{M_{S}^{\prime}}^{2}\right)^{2}+8(a_{H}+a_{H}^{\prime})^{2} v^{2}~,\nonumber\\
D&=v^{2}\left(\lambda_{DH}-\lambda_{DH}^{\prime}+2 \lambda_{DH}^{\prime \prime}-\lambda_{SH}-2 \lambda_{SH}^{\prime}\right)\left(4\left(M_{D}^{2}-M_{S}^{2}-2 {M_{S}^{\prime}}^{2}\right)\right.\nonumber\\
&\left.+v^{2}\left(\lambda_{DH}-\lambda_{DH}^{\prime}+2 \lambda_{DH}^{\prime \prime}-\lambda_{SH}-2 \lambda_{SH}^{\prime}\right)\right)~.
    \end{split}
\end{equation}

In addition, the couplings between the vectorlike quark $Q^{\prime}$ and the up/down-type quarks 
are the same as~\cite{Calibbi:2019bay,Li:2022eby}. And
 the lightest real or imaginary part of the scalars is the DM candidate.

\section{Flavour Observables and Dark Matter}\label{Section3}

In the SESM, besides the two charged scalar particles, there are four neutral scalar particles, the real (imaginary) parts of scalar fields 
$S_{1}$($S_{1}^{\prime}$) and $S_{2}$($S_{2}^{\prime}$). In the following, we consider either $S_{1}$ or $S_{1}^{\prime}$ be the DM candidate  and assume the mass difference between $S_{1}$ and $S_{1}^{\prime}$ is less than several GeVs, and then
 they can be regarded as approximately degenerate.

\subsection{$R_{K}$ and $B_{s}$ mixing}

The effective operators contributing to $R_{K}$  are~\cite{Calibbi:2019bay}
\begin{equation}
\mathcal{H}_{\mathrm{eff}}^{b s \mu \mu} \supset-\mathcal{N}\left[C_{9}^{b s \mu \mu}\left(\bar{s} \gamma_{\mu} P_{L} b\right)\left(\bar{\mu} \gamma^{\mu} \mu\right)+C_{10}^{b s \mu \mu}\left(\bar{s} \gamma_{\mu} P_{L} b\right)\left(\bar{\mu} \gamma^{\mu} \gamma_{5} \mu\right)+\text { h.c. }\right]~,
\end{equation}
with the normalization factor
\begin{equation}
\mathcal{N} \equiv \frac{4 G_{F}}{\sqrt{2}} \frac{e^{2}}{16 \pi^{2}} V_{t b} V_{t s}^{*}~.
\end{equation}
The contributions to $C_{9,10}^{bs\mu\mu}$ induced by the exotic particles can be written as
\begin{align}
&\Delta C_{9}^{b s \mu \mu}=-\frac{\lambda_{3}^{Q_{d}} \lambda_{2}^{Q_{d} *}}{128 \pi^{2} \mathcal{N}} \sum_{\alpha=1,2} \frac{\left|U_{1 \alpha}\right|^{4}\left|\lambda_{2}^{L}\right|^{2}+\left|U_{1 \alpha}\right|^{2}\left|U_{2 \alpha}\right|^{2}\left|\lambda_{2}^{E}\right|^{2}}{M_{S_{\alpha}}^{2}} F_{2}\left(\frac{M_{Q}^{2}}{M_{S_{\alpha}}^{2}}, \frac{M_{L}^{2}}{M_{S_{\alpha}}^{2}}\right)~,\label{19}\\
&\Delta C_{10}^{b s \mu \mu}=\frac{\lambda_{3}^{Q_{d}} \lambda_{2}^{Q_{d}} *}{128 \pi^{2} \mathcal{N}} \sum_{\alpha=1,2} \frac{\left|U_{1 \alpha}\right|^{4}\left|\lambda_{2}^{L}\right|^{2}-\left|U_{1 \alpha}\right|^{2}\left|U_{2 \alpha}\right|^{2}\left|\lambda_{2}^{E}\right|^{2}}{M_{S_{\alpha}}^{2}} F_{2}\left(\frac{M_{Q}^{2}}{M_{S_{\alpha}}^{2}}, \frac{M_{L}^{2}}{M_{S_{\alpha}}^{2}}\right)~,
\label{20}
\end{align}
with the loop function
\begin{equation}
F_{2}(x, y) \equiv \frac{1}{(x-1)(y-1)}+\frac{x^{2} \log x}{(x-1)^{2}(x-y)}+\frac{y^{2} \log y}{(y-1)^{2}(y-x)}~.
\end{equation}

According to the up-to-date fitting \cite{Altmannshofer:2021qrr} (2 $\sigma$), $\Delta C_{9}^{b s \mu \mu}$ and $\Delta C_{10}^{b s \mu \mu}$ need to satisfy
\begin{equation}
    6.74 + 9.04 (\Delta C_{9}^{b s \mu \mu})^2 + \Delta C_{9}^{b s \mu \mu} (14.96 - 10.68 \Delta C_{10}^{b s \mu \mu}) + \Delta C_{10}^{b s \mu \mu} (-13.22 + 11.90 \Delta C_{10}^{b s \mu \mu}) \leq 1~.
\end{equation}
The effective operators contributing to the $B_{s}-\bar{B}_{s}$ oscillations are
\begin{equation}
\mathcal{H}_{\mathrm{eff}}^{b d_{i}} \supset C_{1}^{b d_{i}}\left(\overline{d_{i}} \gamma_{\mu} P_{L} b\right)\left(\overline{d_{i}} \gamma^{\mu} P_{L} b\right)+\text { h.c. } \qquad d_{i}=d,s~.
\end{equation}
The $Q^{\prime}-\Phi_{S}$ box diagram gives Wilson coefficients
\begin{equation}
\Delta C_{1}^{b d_{i}}=\frac{\left(\lambda_{3}^{Q_{d}} \lambda_{i}^{Q_{d}{ }^{*}}\right)^{2}}{128 \pi^{2}} \sum_{\alpha=1,2} \frac{\left|U_{1 \alpha}\right|^{4}}{M_{S_{\alpha}}^{2}} F\left(\frac{M_{Q}^{2}}{M_{S_{\alpha}}^{2}}\right)~,
\end{equation}
with the loop function 
\begin{equation}
F(x) \equiv \frac{x^{2}-1-2 x \log x}{(x-1)^{3}}~.
\end{equation}
The latest bound \cite{DiLuzio:2019jyq,Silvestrini:2018dos} is
\begin{equation}
\Delta C_{1}^{b s}<2.1 \times 10^{-5} \mathrm{TeV}^{-2}~.
\end{equation}

\subsection{Muon Anomalous Magnetic Moment}

The chirally-enhanced contribution induced by exotic particles to  $a_{\mu}$ is
\begin{equation}
\Delta a_{\mu} \approx-\frac{m_{\mu} M_{L}}{8 \pi^{2}} \sum_{\alpha=1,2} \frac{\operatorname{Re}\left(\lambda_{2}^{L} \lambda_{2}^{E *} U_{1 \alpha} U_{2 \alpha}^{*}\right)}{M_{S \alpha}^{2}} f_{L R}\left(\frac{M_{L}^{2}}{M_{S_{\alpha}}^{2}}\right)~,
\end{equation}
where the loop function is
\begin{equation}
f_{L R}(x) \equiv \frac{3-4 x+x^{2}+2 \log x}{2(x-1)^{3}}~.
\end{equation}
The latest $(1\sigma)$ discrepancy is given by \cite{Muong-2:2021ojo}
\begin{equation}
a_{\mu}^{\mathrm{EXP}}-a_{\mu}^{\mathrm{SM}}=(2.51 \pm 0.59) \times 10^{-9}~.
\end{equation}

\subsection{Dark Matter}

With the mass splittings between the real and imaginary parts of new scalar fields, 
the DM co-annihilation process via $Z$ boson appears. In this Section, we mainly consider the $Z$ pole, 
and thus impose the following three conditions for it. First, because the $Z$ boson couples exclusively 
to the CP-even and CP-odd components of $S_{s}^{0}$ and $S_{d}^{0}$,  we should have the mass splitting 
between the real and imaginary parts which can be induced 
by the $a_{H}$, $a_{H}^{\prime}$ ${M_{S}^{\prime}}^{2}$, $\lambda_{SH}^{\prime}$ and $\lambda_{DH}^{\prime\prime}$ terms in Eq.(\ref{lagrangian}). Second, the mass difference between the real and imaginary parts must be small 
in order to realize the co-annihilation process. Third, the doublet component of the DM should be considerable.
\begin{figure}[ht]
    \centering
    \includegraphics[width=0.6\textwidth]{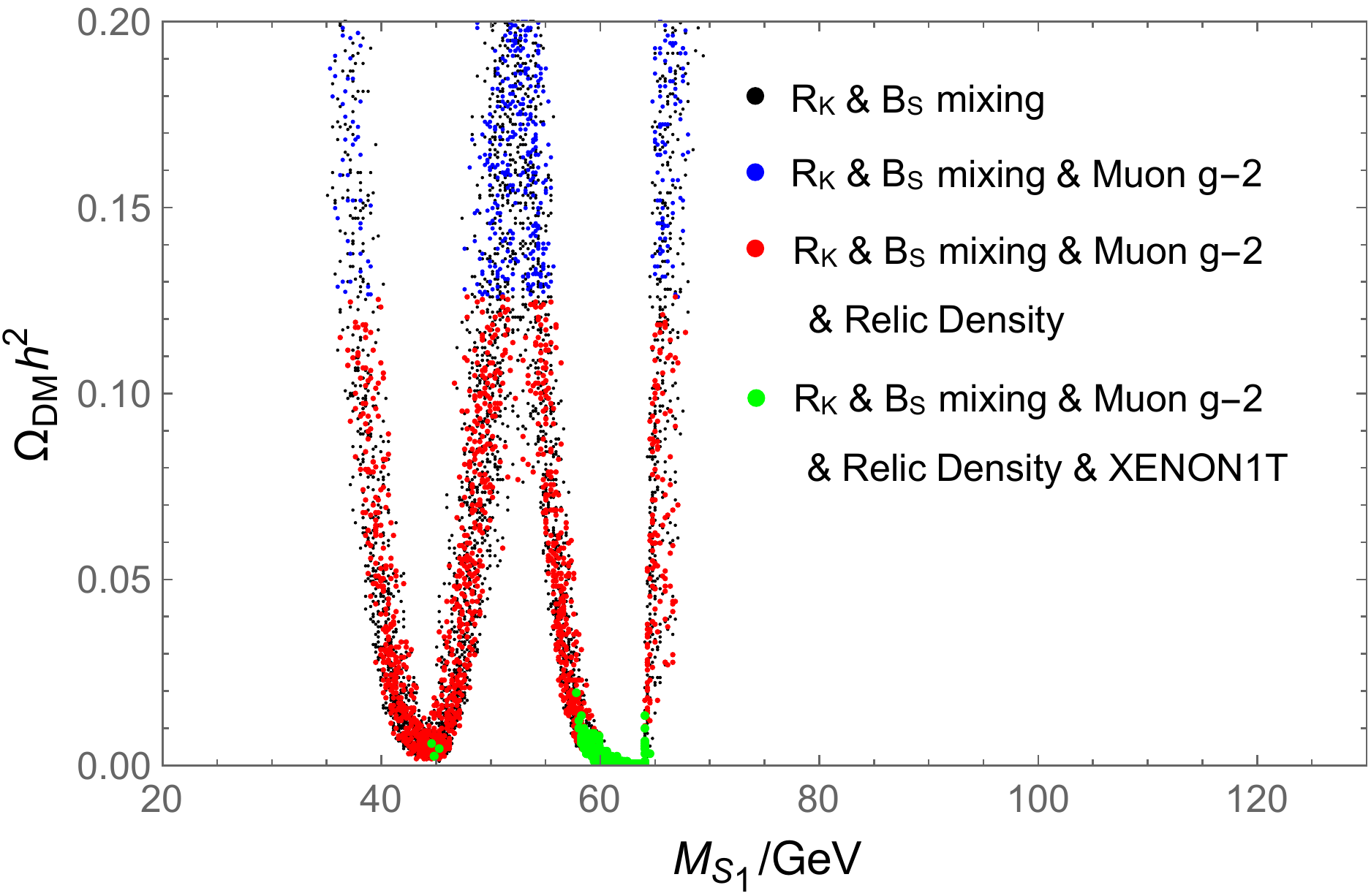}
    \caption{DM (co-)annihilation mediated by $Z$ and Higgs resonances. 
    The black, blue, red, and green dots satisfy the constraints of $R_{K}$ (2 $\sigma$) and $B_{s}-\bar{B}_{s}$, muon g-2 (1 $\sigma$), DM relic density, and XENON1T direct detection in order.
    }
    \label{Z-Higgs-pole}
\end{figure}

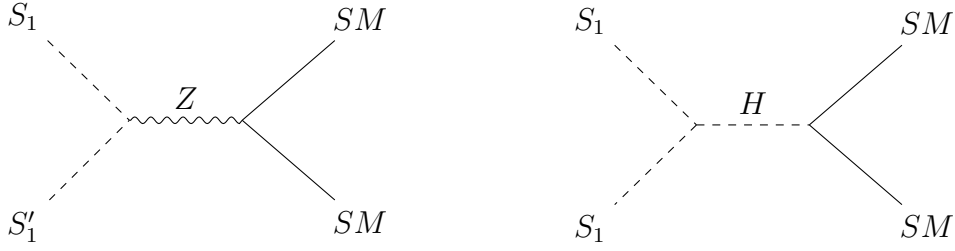
\begin{figure}[ht]
    \centering
 \begin{subfigure}{.45\textwidth}
  \begin{tikzpicture}
  \begin{feynman}
  \vertex (a);
  \vertex [above left=of a](b) {\(S_{1}\)};
  \vertex [below left=of a](c) {\(S_{1}^{\prime}\)};
  \vertex [right=of a](d);
  \vertex [above right=of d](e) {\(SM\)};
  \vertex [below right=of d](f) {\(SM\)};
  
  \diagram*{
  (b) -- [scalar] (a),
  (c) -- [scalar] (a),
  (a) -- [boson, edge label=\(Z\)] (d),
  (e) -- [plain] (d),
  (d) -- [plain] (f),
  };
  \end{feynman}
\end{tikzpicture}
\end{subfigure}
\begin{subfigure}{.45\textwidth}
    \begin{tikzpicture}
  \begin{feynman}
  \vertex (a);
  \vertex [above left=of a](b) {\(S_{1}\)};
  \vertex [below left=of a](c) {\(S_{1}\)};
  \vertex [right=of a](d);
  \vertex [above right=of d](e) {\(SM\)};
  \vertex [below right=of d](f) {\(SM\)};
  
  \diagram*{
  (b) -- [scalar] (a) -- [scalar] (c),
  (a) -- [scalar, edge label=\(H\)] (d),
  (e) -- [plain] (d),
  (d) -- [plain] (f),
  };
  \end{feynman}
  \end{tikzpicture}
    \end{subfigure}
    \caption{The DM (co-)annihilation process.}
    \label{DM-annihilation}
\end{figure}

In Fig.~\ref{Z-Higgs-pole}, we illustrate the points satisfied the constraints of $R_{K}$ (2 $\sigma$) and $B_{s}-\bar{B_{s}}$ (black points), the muon g-2 (1 $\sigma$, blue points), DM relic density (red points) and XENON1T direct detection (green points). We can obtain the correct dark matter relice density
via the Higgs mediated DM annihilation as before, and we do not need to consider the inelastic dark matter 
in this case. Interestingly, we can also obtain the correct dark matter relice density via
the $Z$ mediated  co-annihilation, which is a new feature compared to Ref.~\cite{Li:2022eby}.
We need to point out that the direct detection is so strict in the small mass regions that the cross section should be rescaled, thus the direct detection favors the undersaturated DM abundance. 
The corresponding DM (co-)annihilation processes are demonstrated in Fig.~\ref{DM-annihilation},
and the benchmark points (some of the green points in Fig.~\ref{Z-Higgs-pole}) are displayed 
in Table~\ref{Benchmark points}. Point 1 and Point 2 correspond to the $Z$-mediated DM co-annihilation processes, 
Point 3 and Point 4 correspond to the Higgs-mediated DM annihilation processes. All these points 
satisfied the constraints of $R_{K}$, $B_{s}-\bar{B_{s}}$, muon g-2, DM relic density, and XENON1T direct detection simultaneously.

Finally, we would like to comment about the electroweak precision. Since first we impose the $Z_{2}$ symmetry in the model, where SM particles are $Z_{2}$ even and the new particles are $Z_{2}$ odd. Second, the mass splittings between charged parts and neutral parts of scalar and new fermions are assumed to be zero, thus there is no limitation to electroweak precision tests.

\begin{table}[ht]
    \centering
    \begin{tabular}{|c|cccc|}
    \hline
         &Point 1 &Point 2 &Point 3 &Point 4 \\
    \hline
           $\text{Relic density}$&0.002324&0.004407&0.000056&0.000103 \\
          $M_{S_{1}}\text{/GeV}$&44.9020863& 45.1852846&62.6676439&62.8656877 \\
          $M_{S_{2}}\text{/GeV}$&105.757632&105.721226&117.245807&129.928019\\
          $M_{S_{1}^{\prime}}\text{/GeV}$&44.568636&44.8051653&63.3885604&62.9503338\\
          $M_{S_{2}^{\prime}}\text{/GeV}$&105.879782&105.816872&116.41058&129.869903\\
          $a_{H}\text{/GeV}$&-16.5052592&-14.3744695&-19.73936&-18.3329214\\
          $a_{H}^{\prime}\text{/GeV}$&0.13992708&0.0582310513&0.57179383&0.0376930898\\
          $M_{S}\text{/GeV}$&54.9009796&52.3632109&73.3372329&69.3362989\\
          $M_{S}^{\prime}\text{/GeV}$&0.528484022&1.9419098&2.52399856&3.61952121\\
          $M_{D}\text{/GeV}$&101.033086&101.974629&110.642889&126.70021\\
          $M_{L^{\prime}}\text{/GeV}$&729.517764&742.893444&854.330313&942.259851\\
          $M_{Q^{\prime}}\text{/GeV}$&3902.10638&3937.67224&3925.43003&3904.98369\\
          $\lambda_{SH}$&-0.000133829454&0.000882003049&0.000915186305&-0.000205331286\\
          $\lambda_{SH}^{\prime}$&-0.000641529771& -0.000559134158&-0.00062157876&-0.000389952102\\
          $\lambda_{DH}$&-0.000593605144&0.000790845034&-0.000583304638&-0.0000846866652\\
          $\lambda_{DH}^{\prime}$&0.0000286857473&-0.000668477037&0.000283245864&0.000676767623\\
          $\lambda_{DH}^{\prime\prime}$&0.000599468449&0.000319470619&-0.000448653707&-0.0000789446352\\
          $\lambda_{2}^{Q}$&-0.477980508&-0.419005841&-0.441268492&-0.437252366 \\
          $\lambda_{3}^{Q}$&0.842752986&0.859113671&0.756431246&0.837793862\\
          $\lambda_{2}^{L}$&2.23637283&2.47117928&2.48490618&2.22319786\\
          $\lambda_{2}^{E}$&0.00183980931&0.00243037424&0.00156155801&0.00260726908\\
          $\lambda_{S}/2$&-0.129885309&-0.186246721&-0.190968994&-0.248878009\\
          $\lambda_{S}^{\prime}/2$&-0.763678112&-0.674083096&-0.893186053&-0.760477119\\
          $\lambda_{S}^{\prime\prime}/2$&-0.688219001&-0.522936927&-0.581113359&-0.516623688\\
          $\lambda_{D}/2$&-0.652102746& -0.797238171&-0.626557022&-0.787219103\\
          $\lambda_{SD}$&1.7597484&1.71116141&1.83634899&1.70180638\\
          $\lambda_{SD}^{\prime}$&0.862490095&0.741513741&0.702268647&0.861440253\\
          
    \hline
    \end{tabular}
    \captionsetup{justification=raggedright}
    \caption{The benchmark points for the DM annihilation and co-annihilation. All the points are consistent with $R_{K}$, $B_{s}-\bar{B}_{s}$, muon g-2, Xenon1T direct detection, and DM relic density simultaneously.}
    \label{Benchmark points}
\end{table}

\section{The CP Violation and Inelastic Dark Matter} \label{Section4}

In this Section, we shall consider the CP-violation and inelastic dark matter.
The CP-violation can be realized in our model by considering the complex couplings 
of $a_{H}$, $a_{H}^{\prime}$, and $\lambda_{SH}^{\prime}$ in Eq.~\ref{lagrangian}. The modified mass matrix is given by
\begin{equation}
\setlength{\arraycolsep}{0pt}
\tiny{
    \left(\begin{array}{cccc}
{M_{S}}^{2}+2 {M_{S}^{\prime}}^{2}+\frac{v^{2} \lambda_{SH}}{2}+v^{2} Re(\lambda_{SH}^{\prime}) &-Im(\lambda_{SH}^{\prime}) v^{2}& \frac{\left(Re(a_{H}^{\prime})+Re(a_{H})\right)v}{\sqrt{2}}&\frac{\left(Im(a_{H})+Im(a_{H}^{\prime})\right)v}{\sqrt{2}}\\[6pt]
-Im(\lambda_{SH}^{\prime}) v^{2}&{M_{S}}^{2}-2 {M_{S}^{\prime}}^{2}+\frac{v^{2} \lambda_{SH}}{2}-v^{2} Re(\lambda_{SH}^{\prime}) &\frac{\left(Im(a_{H})-Im(a_{H}^{\prime})\right)v}{\sqrt{2}}&\frac{\left(Re(a_{H}^{\prime})-Re(a_{H})\right)v}{\sqrt{2}}\\[6pt]
\frac{\left(Re(a_{H}^{\prime})+Re(a_{H})\right)v}{\sqrt{2}}&\frac{\left(Im(a_{H})-Im(a_{H}^{\prime})\right)v}{\sqrt{2}}&M_{D}^{2}+\frac{v^{2} \lambda_{DH}}{2}-\frac{v^{2} \lambda_{DH}^{\prime}}{2}+v^{2} \lambda_{DH}^{\prime \prime}&0\\[6pt]
\frac{\left(Im(a_{H})+Im(a_{H}^{\prime})\right)v}{\sqrt{2}}&\frac{\left(Re(a_{H}^{\prime})-Re(a_{H})\right)v}{\sqrt{2}}&0&M_{D}^{2}+\frac{v^{2} \lambda_{DH}}{2}-\frac{v^{2} \lambda_{DH}^{\prime}}{2}-v^{2} \lambda_{DH}^{\prime \prime}\\
\end{array}\right)}~.
\label{RIMixing}
\end{equation}

From Eq.~\ref{RIMixing}, we obtain the mass matrices in Eq~\ref{Real} and Eq~\ref{Imaginary} 
when we turn off the imaginary parts of $a_{H}$, $a_{H}^{\prime}$, and $\lambda_{SH}^{\prime}$. 
The general interactions between the new scalars and SM Higgs field are given in~\ref{A3}, 
which corresponds to scenario III. In scenario I, II, and IV, we have two block diagonal mass matrixes 
since the absence of imaginary parts of coupling constants. Then the interactions can be 
reduced to ~\ref{A1}, ~\ref{A2} and ~\ref{A4}.
The mass eigenstate are numerically calculated in the following discussions. 

These complex couplings can induce the inelastic DM scattering processes, and
the  $S_{1} S_{1}^{\prime}h$ coupling arises due to the CP-violation interaction between exotic scalar fields 
and SM Higgs boson. In this Section, we investigate four kinds of inelastic DM scenarios and there exist the viable parameter space in our model. For simplicity, we give the benchmark points which are consistent with $R_{K}$, $B_{s}-\bar{B}_{s}$, muon g-2, Xenon1T direct detection, and DM relic density simultaneously. Moreover, the Higgs mediated inelastic processes 
can saturate the DM relic density while $Z$ mediated one is undersaturated 
due to the smaller scalar mass splitting.

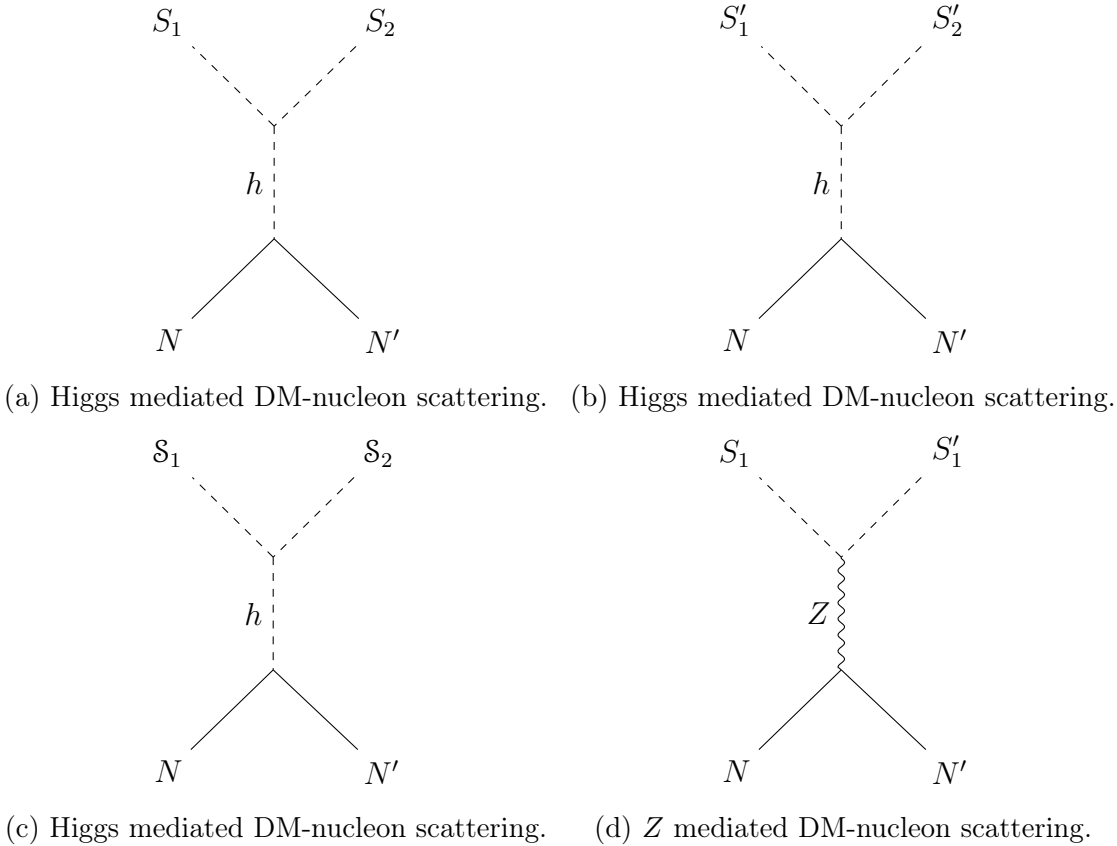
\begin{figure}[h]
    \begin{subfigure}{0.45\textwidth}
        \centering
  \begin{tikzpicture}
       \begin{feynman}
       \vertex (m1);
       \vertex [above left=of m1](l1) {\(S_{1}\)};
       \vertex [above right=of m1](r1) {\(S_{2}\)};
       \vertex [below=of m1](m2);
       \vertex [below left=of m2](l2) {\(N\)};
       \vertex [below right=of m2](r2) {\(N^{\prime}\)};
  
       \diagram*{
  (r1) -- [scalar] (m1) -- [scalar] (l1),
  (m1) -- [scalar, edge label'=\(h\)] (m2),
  (l2) -- [plain] (m2) -- [plain] (r2),
  };
       \end{feynman}
       \end{tikzpicture}
    \caption{Higgs mediated DM-nucleon scattering.}
    \label{InelasticDMa}
    \end{subfigure}
\begin{subfigure}{0.45\textwidth}
        \centering
  \begin{tikzpicture}
       \begin{feynman}
       \vertex (m1);
       \vertex [above left=of m1](l1) {\(S_{1}^{\prime}\)};
       \vertex [above right=of m1](r1) {\(S_{2}^{\prime}\)};
       \vertex [below=of m1](m2);
       \vertex [below left=of m2](l2) {\(N\)};
       \vertex [below right=of m2](r2) {\(N^{\prime}\)};
  
       \diagram*{
  (r1) -- [scalar] (m1) -- [scalar] (l1),
  (m1) -- [scalar, edge label'=\(h\)] (m2),
  (l2) -- [plain] (m2) -- [plain] (r2),
  };
       \end{feynman}
       \end{tikzpicture}
    \caption{Higgs mediated DM-nucleon scattering.}
    \label{InelasticDMb}
\end{subfigure}
\\
    \begin{subfigure}{0.45\textwidth}
        \centering
  \begin{tikzpicture}
       \begin{feynman}
       \vertex (m1);
       \vertex [above left=of m1](l1) {\(\mathscr{S}_{1}\)};
       \vertex [above right=of m1](r1) {\(\mathscr{S}_{2}\)};
       \vertex [below=of m1](m2);
       \vertex [below left=of m2](l2) {\(N\)};
       \vertex [below right=of m2](r2) {\(N^{\prime}\)};
  
       \diagram*{
  (r1) -- [scalar] (m1) -- [scalar] (l1),
  (m1) -- [scalar, edge label'=\(h\)] (m2),
  (l2) -- [plain] (m2) -- [plain] (r2),
  };
       \end{feynman}
       \end{tikzpicture}
    \caption{Higgs mediated DM-nucleon scattering.}
    \label{InelasticDMc}
    \end{subfigure}
\begin{subfigure}{0.45\textwidth}
        \centering
  \begin{tikzpicture}
       \begin{feynman}
       \vertex (m1);
       \vertex [above left=of m1](l1) {\(S_{1}\)};
       \vertex [above right=of m1](r1) {\(S_{1}^{\prime}\)};
       \vertex [below=of m1](m2);
       \vertex [below left=of m2](l2) {\(N\)};
       \vertex [below right=of m2](r2) {\(N^{\prime}\)};
  
       \diagram*{
  (r1) -- [scalar] (m1) -- [scalar] (l1),
  (m1) -- [boson, edge label'=\(Z\)] (m2),
  (l2) -- [plain] (m2) -- [plain] (r2),
  };
       \end{feynman}
       \end{tikzpicture}
    \caption{$Z$ mediated DM-nucleon scattering.}
    \label{InelasticDMd}
\end{subfigure}
\caption{The inelastic DM-nucleon scattering processes. }
\end{figure}

\subsection{Scenario I: Higgs mediated $S_{1} N \rightarrow S_{2} N^{\prime}$}

The Higgs mediated inelastic DM-nucleon scattering process is shown in Fig.~\ref{InelasticDMa}, 
where $S_{1}$ and $S_{2}$ are the real parts of scalar fields, $N$ and $N^{\prime}$ are nucleons before and after scattering. To achieve this kind of scenario, 
we need to suppress the couplings like $S_{1} S_{1} h$, $S_{1} S_{1}^{\prime} h$, and $S_{1} S_{2}^{\prime} h$, etc. 

In Table~\ref{scenarioI}, we elaborate a concrete realization of scenario I,
 and the corresponding benchmark point is Point 1 in Table~\ref{Benchmark points Inelastic}, where all the points are consistent with $R_{K}$, $B_{s}-\bar{B}_{s}$, muon g-2, Xenon1T direct detection, and DM relic density simultaneously.
We denote the red marks as the most efficient vertex to realize the corresponding scenario,
 as well as employ the blue and green marks as the vertexes which are forbidden by the mixing angle 
and mass splitting, respectively. 
For instance, we choose $\lambda_{DH}^{\prime \prime}=0.05$, $\lambda_{DH}=0.2$, $\lambda_{DH}^{\prime}=0.1$, $Re(a_{H})=-Re(a_{H}^{\prime})=0.138149275$, and $\lambda_{SH}=Re(\lambda_{SH}^{\prime})=Im(\lambda_{SH}^{\prime})=Im(a_{H})=Im(a_{H}^{\prime})=0$. 
One can check that the opposite sign between $a_{H}$ and $a_{H}^{\prime}$  eliminates the mixing 
of imaginary parts of the singlet and doublet scalars, 
which means $U_{12}^{\prime}=U_{21}^{\prime}=0$, and thus explains the origin of the blue marks in Table~\ref{scenarioI}. 
The $S_{1}S_{2}h$ vertex marked as red is enhanced by the mixing angle of 
the real parts of the singlet and doublet scalars, 
and is responsible for  efficiently realizing this kind of scenario. 
All the remaining vertexes exactly vanish by a comprehensive consideration 
of the parameters we take above.

\begin{table}[ht]
    \centering
    \begin{tabular}{c|c|c|c|c|c|c|c}
         &couplings&  $S_{1}S_{1}h$&$S_{1}S_{1}^{\prime}h$&$S_{1}S_{2}h$&$S_{1} S_{2}^{\prime} h$&$S_{1}^{\prime} S_{1}^{\prime} h$&$S_{1}^{\prime} S_{2}^{\prime} h$\\
         \hline
         \multirow{4}{*}{$S_{i}S_{j}h$}
         &$(2\lambda_{DH}^{\prime\prime}-\lambda_{DH}+\lambda_{DH}^{\prime})U_{i2}U_{j2}$&\ding{55}&&\ding{55}&&&\\
         &$\left(Re(a_{H})-Re(a_{H}^{\prime})\right)U_{i2}U_{j1}$&\ding{51}&&\ding{51}&&&\\
         &$\left(Re(a_{H})-Re(a_{H}^{\prime})\right)U_{i1}U_{j2}$&\ding{51}&&\textcolor{red}{\ding{51}}&&&\\
         &$\left(2Re(\lambda_{SH}^{\prime})-\lambda_{SH}\right)U_{i1}U_{j1}$&\ding{55}&&\ding{55}&&&\\
         \hline
         \multirow{4}{*}{$S_{i}^{\prime}S_{j}^{\prime}h$}
         &$(2\lambda_{DH}^{\prime\prime}+\lambda_{DH}-\lambda_{DH}^{\prime})U_{i2}^{\prime}U_{j2}^{\prime}$&&&&&\textcolor{blue}{\ding{55}}&\textcolor{blue}{\ding{55}}\\
         &$\left(Re(a_{H})+Re(a_{H}^{\prime})\right)U_{i1}^{\prime}U_{j2}^{\prime}$&&&&&\ding{55}&\ding{55}\\
         &$\left(Re(a_{H})+Re(a_{H}^{\prime})\right)U_{i2}^{\prime}U_{j1}^{\prime}$&&&&&\ding{55}&\ding{55}\\         &$\left(2Re(\lambda_{SH}^{\prime})+\lambda_{SH}\right)U_{i1}^{\prime}U_{j1}^{\prime}$&&&&&\ding{55}&\ding{55}\\
         \hline
         \multirow{4}{*}{$S_{i}S_{j}^{\prime}h$}
         &$\left(Im(a_{H}^{\prime})-Im(a_{H})\right)U_{i2}U_{j1}^{\prime}$&&\ding{55}&&\ding{55}&&\\
         &$\left(Im(a_{H})+Im(a_{H}^{\prime})\right)U_{i1}U_{j2}^{\prime}$&&\ding{55}&&\ding{55}&&\\
         &$Im(\lambda_{SH}^{\prime})U_{i1}^{\prime}U_{j1}^{\prime}$&&\ding{55}&&\ding{55}&&   \\
         \hline
    \end{tabular}
    \caption{The couplings in scenario I.}
    \label{scenarioI}
\end{table}

\subsection{Scenario II: Higgs mediated $S_{1}^{\prime} N \rightarrow S_{2}^{\prime} N^{\prime}$}

 As shown in Fig.~\ref{InelasticDMb}, we come to the second scenario, 
in which the DM candidate is the imaginary part of scalar field. 
The concrete realization is presented in Table~\ref{scenarioII}, and the benchmark point is Point 2 
in Table~\ref{Benchmark points Inelastic}. Similar to scenario I, the same sign between $a_{H}$ and $a_{H}^{\prime}$  
eliminates the mixing of real parts of the singlet and doublet scalars, 
which means $U_{12}=U_{21}=0$. The corresponding vertexes are marked as blue,
 and the red mark is the most efficient vertex to realize scenario II.   

\begin{table}[ht]
    \centering
    \begin{tabular}{c|c|c|c|c|c|c|c}
         &couplings&  $S_{1}S_{1}h$&$S_{1}S_{1}^{\prime}h$&$S_{1}S_{2}h$&$S_{1} S_{2}^{\prime} h$&$S_{1}^{\prime} S_{1}^{\prime} h$&$S_{1}^{\prime} S_{2}^{\prime} h$\\
         \hline
         \multirow{4}{*}{$S_{i}S_{j}h$}
         &$(2\lambda_{DH}^{\prime\prime}-\lambda_{DH}+\lambda_{DH}^{\prime})U_{i2}U_{j2}$&\textcolor{blue}{\ding{55}}&&\textcolor{blue}{\ding{55}}&&&\\
         &$\left(Re(a_{H})-Re(a_{H}^{\prime})\right)U_{i2}U_{j1}$&\ding{55}&&\ding{55}&&&\\
         &$\left(Re(a_{H})-Re(a_{H}^{\prime})\right)U_{i1}U_{j2}$&\ding{55}&&\ding{55}&&&\\
         &$\left(2Re(\lambda_{SH}^{\prime})-\lambda_{SH}\right)U_{i1}U_{j1}$&\ding{55}&&\ding{55}&&&\\
         \hline
         \multirow{4}{*}{$S_{i}^{\prime}S_{j}^{\prime}h$}
         &$(2\lambda_{DH}^{\prime\prime}+\lambda_{DH}-\lambda_{DH}^{\prime})U_{i2}^{\prime}U_{j2}^{\prime}$&&&&&\ding{55}&\ding{55}\\
         &$\left(Re(a_{H})+Re(a_{H}^{\prime})\right)U_{i1}^{\prime}U_{j2}^{\prime}$&&&&&\ding{51}&\textcolor{red}{\ding{51}}\\
         &$\left(Re(a_{H})+Re(a_{H}^{\prime})\right)U_{i2}^{\prime}U_{j1}^{\prime}$&&&&&\ding{51}&\ding{51}\\         &$\left(2Re(\lambda_{SH}^{\prime})+\lambda_{SH}\right)U_{i1}^{\prime}U_{j1}^{\prime}$&&&&&\ding{55}&\ding{55}\\
         \hline
         \multirow{4}{*}{$S_{i}S_{j}^{\prime}h$}
         &$\left(Im(a_{H}^{\prime})-Im(a_{H})\right)U_{i2}U_{j1}^{\prime}$&&\ding{55}&&\ding{55}&&\\
         &$\left(Im(a_{H})+Im(a_{H}^{\prime})\right)U_{i1}U_{j2}^{\prime}$&&\ding{55}&&\ding{55}&&\\
         &$Im(\lambda_{SH}^{\prime})U_{i1}^{\prime}U_{j1}^{\prime}$&&\ding{55}&&\ding{55}&&   \\
         \hline
    \end{tabular}
    \caption{The couplings in scenario II.}
    \label{scenarioII}
\end{table}

\subsection{Scenario III: Higgs mediated $\mathscr{S}_{1} N \rightarrow \mathscr{S}_{2} N^{\prime}$}

Since the couplings $a_{H}$, $a_{H}^{\prime}$, and $\lambda_{SH}^{\prime}$ can be complex numbers, 
the CP-violation interactions with Higgs field arise. 
In this scenario, we employ four-dimensional mass matrix as shown in Eq.~\ref{RIMixing},
 and the mass eigenstates are denoted by $\mathscr{S}_i$.
Therefore, we have $\mathscr{S}_{1}\mathscr{S}_{2}h$ interaction as presented in Fig.~\ref{InelasticDMc},
 which shows the coupling between 
that Higgs field, CP-even and CP-odd parts of scalar fields.

In Table~\ref{scenarioIII}, the green marks mean that 
the $\mathscr{S}_{1}\mathscr{S}_{4}h$ and $\mathscr{S}_{2}\mathscr{S}_{4}h$ interactions 
are suppressed by the mass splittings between $\mathscr{S}_{1}$ and $\mathscr{S}_{4}$, as well as
 $\mathscr{S}_{2}$ and $\mathscr{S}_{4}$, respectively. 
The benchmark point is Point 3 in Table~\ref{Benchmark points Inelastic}.

\begin{table}[ht]
    \centering
    \begin{tabular}{c|c|c|c|c|c|c|c}
         &couplings&  $\mathscr{S}_{1}\mathscr{S}_{1}h$&$\mathscr{S}_{1}\mathscr{S}_{2}h$&$\mathscr{S}_{1}\mathscr{S}_{3}h$&$\mathscr{S}_{1} \mathscr{S}_{4} h$&$\mathscr{S}_{2} \mathscr{S}_{2} h$&$\mathscr{S}_{2} \mathscr{S}_{4} h$\\
         \hline
         &$(2\lambda_{DH}^{\prime\prime}-\lambda_{DH}+\lambda_{DH}^{\prime})U_{j2}U_{k2}$&\ding{55}&\ding{55}&\ding{55}&\ding{55}&\ding{55}&\ding{55}\\
         &$\left(Re(a_{H})-Re(a_{H}^{\prime})\right)U_{j1}U_{k2}$&\ding{55}&\ding{55}&\ding{55}&\ding{55}&\ding{55}&\ding{55}\\
         &$\left(Re(a_{H})-Re(a_{H}^{\prime})\right)U_{j2}U_{k1}$&\ding{55}&\ding{55}&\ding{55}&\ding{55}&\ding{55}&\ding{55}\\
         &$\left(2Re(\lambda_{SH}^{\prime})-\lambda_{SH}\right)U_{j1}U_{k1}$&\ding{55}&\ding{55}&\ding{55}&\ding{55}&\ding{55}&\ding{55}\\
        \multirow{4}{*}{$\mathscr{S}_{j}\mathscr{S}_{k}h$}
         &$(2\lambda_{DH}^{\prime\prime}+\lambda_{DH}-\lambda_{DH}^{\prime})U_{j4}U_{k4}$&\ding{55}&\ding{55}&\ding{55}&\ding{55}&\ding{55}&\ding{55}\\
         &$\left(2Re(\lambda_{SH}^{\prime})+\lambda_{SH}\right)U_{j3}U_{k3}$&\ding{55}&\ding{55}&\ding{55}&\ding{55}&\ding{55}&\ding{55}\\
         &$\left(Re(a_{H})+Re(a_{H}^{\prime})\right)U_{j3}U_{k4}$&\textcolor{blue}{\ding{55}}&\ding{51}&\textcolor{green}{\ding{55}}&\textcolor{green}{\ding{55}}&\ding{51}&\textcolor{green}{\ding{55}}\\
         &$\left(Re(a_{H})+Re(a_{H}^{\prime})\right)U_{j4}U_{k3}$&\textcolor{blue}{\ding{55}}&\ding{51}&\textcolor{green}{\ding{55}}&\textcolor{green}{\ding{55}}&\ding{51}&\textcolor{green}{\ding{55}}\\         
         &$\left(Im(a_{H}^{\prime})-Im(a_{H})\right)U_{j3}U_{k2}$&\textcolor{blue}{\ding{55}}&\ding{51}&\textcolor{green}{\ding{55}}&\textcolor{green}{\ding{55}}&\ding{51}&\textcolor{green}{\ding{55}}\\
         &$\left(Im(a_{H}^{\prime})-Im(a_{H})\right)U_{j2}U_{k3}$&\textcolor{blue}{\ding{55}}&\ding{51}&\textcolor{green}{\ding{55}}&\textcolor{green}{\ding{55}}&\ding{51}&\textcolor{green}{\ding{55}}\\
         &$\left(Im(a_{H})+Im(a_{H}^{\prime})\right)U_{j4}U_{k1}$&\ding{55}&\ding{51}&\textcolor{green}{\ding{55}}&\textcolor{green}{\ding{55}}&\textcolor{blue}{\ding{55}}&\textcolor{green}{\ding{55}}\\
         &$\left(Im(a_{H})+Im(a_{H}^{\prime})\right)U_{j1}U_{k4}$&\ding{51}&\ding{51}&\textcolor{green}{\ding{55}}&\textcolor{green}{\ding{55}}&\textcolor{blue}{\ding{55}}&\textcolor{green}{\ding{55}}\\
         &$Im(\lambda_{SH}^{\prime})U_{j3}U_{k1}$&\ding{51}&\ding{51}&\textcolor{green}{\ding{55}}&\textcolor{green}{\ding{55}}&\textcolor{blue}{\ding{55}}&\textcolor{green}{\ding{55}}  \\
         &$Im(\lambda_{SH}^{\prime})U_{j1}U_{k3}$&\ding{51}&\ding{51}&\textcolor{green}{\ding{55}}&\textcolor{green}{\ding{55}}&\textcolor{blue}{\ding{55}}&\textcolor{green}{\ding{55}}   \\
         \hline
    \end{tabular}
    \caption{The couplings in scenario III.}
    \label{scenarioIII}
\end{table}

\subsection{Scenario IV: $Z$ mediated $S_{1} N \rightarrow S_{1}^{\prime} N^{\prime}$}

As shown in Figure~\ref{InelasticDMd}, because there exists the mass splitting between the real and imaginary parts 
of scalar fields, we have the $Z$ mediated DM-nucleon inelastic scattering. 
The $S_{1}S_{1}^{\prime}Z$ interaction will be large when 
the mass difference between $S_{1}$ and $S_{1}^{\prime}$ is small, and the mixings in both real and imaginary parts 
are large. The corresponding benchmark point is Point 4 in Table~\ref{Benchmark points Inelastic}, where we rescaled the cross section of DM direct direction by $\sigma^{\text{Rescaled}}_{SI}= \sigma_{SI} \cdot \Omega_{\text{DM}}/0.12$ .

\begin{table}[htp]
    \centering
    \begin{tabular}{|c|cccc|}
    \hline
         &Point 1 &Point 2 &Point 3 &Point 4 \\
    \hline
           $\text{Relic density}$&0.120245&0.120063&0.121847&0.00273 \\
          $M_{S_{1}}\text{/GeV}$&62.5280891&63&     63.7565163  & 45.2614126\\
          $M_{S_{2}}\text{/GeV}$&64.5715733&101.106889& 145.249  &104.765951\\
          $M_{S_{1}^{\prime}}\text{/GeV}$&63.3955834&62.8787699&64.0200287  &45.2712507\\
          $M_{S_{2}^{\prime}}\text{/GeV}$&101.106889&64.6181886& 145.347221 &104.766473\\
          $\text{Re}(a_{H})\text{/GeV}$&0.138149275&0.161219197&2 &16\\
          $\text{Re}(a_{H}^{\prime})\text{/GeV}$&-0.138149275&0.161219197&2  &0\\
          $\text{Im}(a_{H})\text{/GeV}$&0&0& 0.0334546373 &0\\
          $\text{Im}(a_{H}^{\prime})\text{/GeV}$&0&0& 0.0503546759 &0\\
          $M_{S}\text{/GeV}$&63&63& 64 &55\\
          $M_{S}^{\prime}\text{/GeV}$&5&0& 3.87583889 &0.5\\
          $M_{D}\text{/GeV}$&64.5&64.5&145.249  &100\\
          $M_{L^{\prime}}\text{/GeV}$&923.760249&890.911643& 902.487 &902.908781\\
          $M_{Q^{\prime}}\text{/GeV}$&3935.5024&3937.63785& 3933.748 &3901.27893\\
          $\lambda_{SH}$&0&0& 0 &0\\
          $\text{Re}(\lambda_{SH}^{\prime})$&0&0& 0 &0\\
          $\text{Im}(\lambda_{SH}^{\prime})$&0&0&  0.000106455208 &0\\
          $\lambda_{DH}$&0.2&0.2&0  &0\\
          $\lambda_{DH}^{\prime}$&0.1&0.1& 0 &0\\
          $\lambda_{DH}^{\prime\prime}$&0.05&-0.05&0  &0\\
          $\lambda_{2}^{Q}$&-0.31671602&-0.5704031& -0.822821 &-0.517382525 \\
          $\lambda_{3}^{Q}$&0.880289647&0.68180215& 0.417204 &0.555727792\\
          $\lambda_{2}^{L}$&2.8761966&2.46163782&2.16895  &2.76669117\\
          $\lambda_{2}^{E}$&0.003504851&0.00252737& 0.00534 &0.001891458\\
          $\lambda_{S}/2$&-0.181683181&-0.339379484& 0.36714921 &0.042653427\\
          $\lambda_{S}^{\prime}/2$&0.897108085&-0.71839593& -0.37282701 &-0.661534232\\
          $\lambda_{S}^{\prime\prime}/2$&-0.327292676&0.340974767& -0.467026357 &-0.055752432\\
          $\lambda_{D}/2$&-0.508257921&0.649093836& -0.656446092 &0.853623195\\
          $\lambda_{SD}$&0.032597077&0.806907755& 0.474364027 &0.852904036\\
          $\lambda_{SD}^{\prime}$&-0.461554286&-0.192091407& 0.68062965 &0.825947186\\
          $\sigma^{\text{Rescaled}}_{SI}/\text{cm}^{2}$&$1.139\times 10^{-49}$&$2.97\times 10^{-49}$&$2.4\times10^{-50}$&$4.68\times10^{-47}$\\
          
    \hline
    \end{tabular}
    \captionsetup{justification=raggedright}
    \caption{The benchmark points for inelastic DM. All the points are consistent with $R_{K}$, $B_{s}-\bar{B}_{s}$, muon g-2, Xenon1T direct detection, and DM relic density simultaneously.}
    \label{Benchmark points Inelastic}
\end{table}

\section{The XENONnT Experimental Constraints}\label{Section5}

\begin{figure}[ht]
    \centering
  \begin{tikzpicture}
       \begin{feynman}
       \vertex (m1);
       \vertex [above left=of m1](l1) {\(e\)};
       \vertex [above right=of m1](r1) {\(e\)};
       \vertex [below=of m1](m2);
       \vertex [below left=of m2](l2) {\(S_{1}\)};
       \vertex [below right=of m2](r2) {\(S_{2}\)};
  
       \diagram*{
  (r1) -- [plain, edge label=\(k\)] (m1) -- [plain, edge label=\(p\)] (l1),
  (m1) -- [scalar, edge label'=\(h\)] (m2),
  (l2) -- [scalar, edge label=\(p_{1}\)] (m2) -- [scalar, edge label=\(k_{1}\)] (r2),
  };
       \end{feynman}
       \end{tikzpicture}
    \caption{The Higgs mediated DM-electron scattering}
    \label{Higgs mediation}
\end{figure}
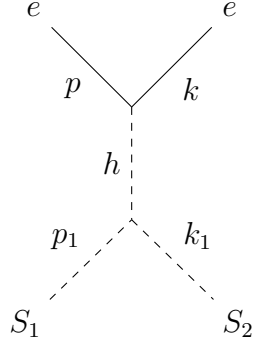

The XENON1T experiment announced the low-energy electronic recoil events below 7 keV, 
which was excluded by the XENONnT experiment. Such latest result sets new limits on
 numerous new physics models. We will investigate the DM-electron inelastic scattering processes 
mediated by Higgs and $Z$ bosons, and discuss the constraints on these two processes 
in light of the XENONnT experiment data. The matrix element of Fig.~\ref{Higgs mediation} is

\begin{equation}
    iM_{\text{free}}=\frac{\bar{u}(k)(-i\frac{y_{e}}{\sqrt{2}})u(p)
\cdot V_{SSh}}{(k_{1}-p_{1})^{2}-M_{h}^{2}}~,
\end{equation}
\begin{equation}
\begin{split}
    V_{SSh}&=-\frac{i}{2}\left(U_{11}\left(\left(2\lambda_{SH}-4\lambda_{SH}^{\prime}\right)v U_{21}+\sqrt{2}\left(a_{H}^{\prime}-a_{H}\right)U_{22}\right)\right.\\
    &\left.+U_{11}\left(\left(2\lambda_{DH}-2\lambda_{DH}^{\prime}-4\lambda_{DH}^{\prime\prime}\right)v U_{22}+\sqrt{2}\left(a_{H}^{\prime}-a_{H}\right)U_{21}\right)
    \right)~,
\end{split}
\end{equation}
where $V_{SSh}$, $U_{ij}$, and $y_{e}$ are the vertex of $S_{1}S_{2} h$, mixing matrix of real part of scalars, 
and Yukawa coupling of electron, respectively.
The differential cross section is 
\begin{equation}
    d \sigma= \frac{1}{2 E_{p} 2 E_{p_1}} \frac{d^{3} \boldsymbol{k}}{(2 \pi)^{3} 2 E_{k}} \frac{d^{3} \boldsymbol{k_{1}}}{(2 \pi)^{3} 2 E_{k_{1}}}(2 \pi)^{4} \delta^{4}\left(p+p_{1}-k-k_{1}\right)\overline{\left|\mathcal{M}_{\text {free }}\right|^{2}}~.
\end{equation}
The real case is that DM scatter with electron in a bound state, 
and thus we should take the wavefunction of initial and final state into account. 
Such effects can be parameterized as an atomic form factor $f_{ion}^{nl}(q)$.

The differential cross section with the fixed DM velocity ($v_{S}$)  is~\cite{Cao:2020bwd,Su:2020zny,Choi:2020ysq}

\begin{equation}
    \frac{d \sigma_{S e} }{d E_{R}}=\frac{\bar{\sigma}_{e}}{8 E_{R} v_{S}^{2} \mu_{S e}^{2}} \int q d q\left|F_{D M}(q)\right|^{2}\sum_{n,l}\left|f_{\text {ion }}^{n l}\left(k, q\right)\right|^{2} ~,
\end{equation}
where $\sigma_{S e}$ is the scattering cross section of DM and a bound electron, 
$E_{R}$ is the recoil energy, $\mu_{Se}$ is the DM-electron reduced mass, $F_{DM}(q)$ is the DM form factor, $f_{\text {ion }}^{n l}\left(k, q\right)$ is the ionization form factor of the $(n,l)$ atomic shell, and $k=\sqrt{2 m_{e} E_{R}}$ is the outgoing momentum of electron.

Following the conventionin Ref.~\cite{Essig:2011nj}, the reference momentum transform 
is fixed at $q=\alpha m_{e}$. And then the matrix element square and  $\bar{\sigma}_{e}$ are defined by
\begin{align}
\overline{\left|\mathcal{M}_{\text {free }}\right|^{2}} &=\overline{\left|\mathcal{M}_{\text {free }}\left(\alpha m_{e}\right)\right|^{2}} \times\left|F_{D M}(q)\right|^{2}, \\
\bar{\sigma}_{e} &=\frac{\mu_{S e}^{2} \overline{\left|\mathcal{M}_{\text {free }}\left(\alpha m_{e}\right)\right|^{2}}}{16 \pi m_{S}^{2} m_{e}^{2}}~.       \end{align}
The event rate can be writen as
\begin{equation}
    \frac{d R}{d E_{R}}=\epsilon\left(E_{R}\right) n_{T} \int_{E_{S}^{\min }}^{E_{S}^{\max }} \frac{d \phi_{S}}{d E_{S}} \frac{d \sigma_{S e}}{d E_{R}}~,
\end{equation}
where $\frac{d\Phi_{S}}{dE_{S}}$ is the DM flux in the Galactic halo, $n_{T}=4.2\times10^{27}$/tonne for Xenon,  and $\epsilon(E_{R})$ is the detection efficiency.
\begin{figure}[ht]
    \centering
    \includegraphics[width=0.9\textwidth]{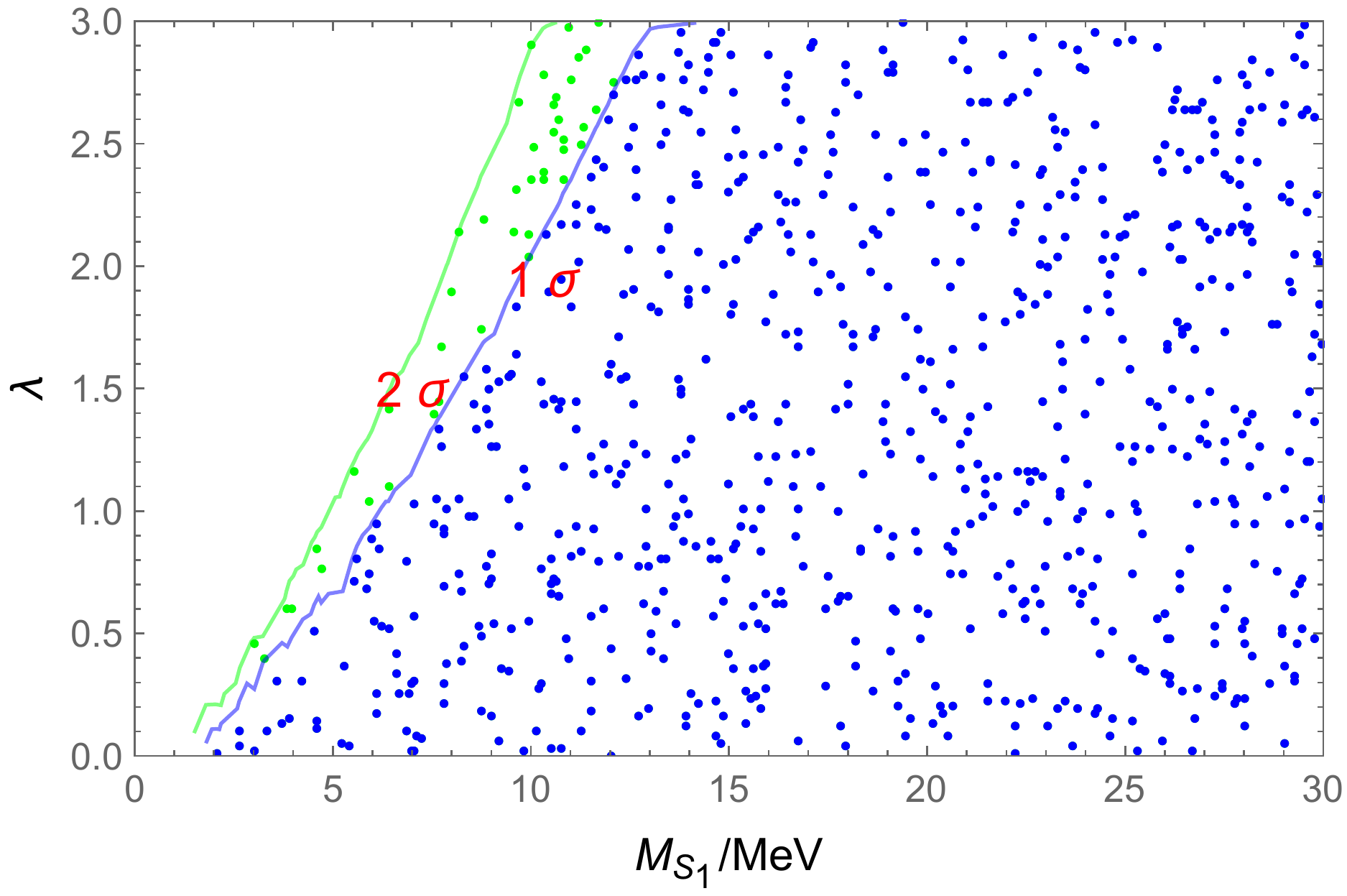}
    \caption{The constraint on the DM-electron scattering process mediated by Higgs particle.}
    \label{XENONnT}
\end{figure}

In Fig.~\ref{XENONnT}, we redefine an effective coupling $\lambda=V_{SSh}/v$ for the small mixing of 
singlet and doublet scalars. A conservative range $\lambda\leq3$ is employed to avoid large quantum corrections,
 and the mass difference between $S_{1}$ and $S_{2}$ is fixed at about 2.5 keV. We present 
the viable parameter space in the $\lambda$ vs $M_{S_1}$ plane for 
the DM-electron scattering process mediated by Higgs particle
 which are consistent with the XENONnT data within 1 $\sigma$ (blue dots) region and 2 $\sigma$ region (blue and green dots). And the DM mass around several MeVs can be excluded (blank region).

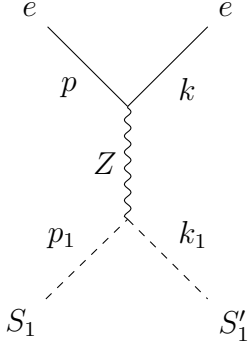
\begin{figure}
    \centering
  \begin{tikzpicture}
       \begin{feynman}
       \vertex (m1);
       \vertex [above left=of m1](l1) {\(e\)};
       \vertex [above right=of m1](r1) {\(e\)};
       \vertex [below=of m1](m2);
       \vertex [below left=of m2](l2) {\(S_{1}\)};
       \vertex [below right=of m2](r2) {\(S_{1}^{\prime}\)};
  
       \diagram*{
  (r1) -- [plain, edge label=\(k\)] (m1) -- [plain, edge label=\(p\)] (l1),
  (m1) -- [boson, edge label'=\(Z\)] (m2),
  (l2) -- [scalar, edge label=\(p_{1}\)] (m2) -- [scalar, edge label=\(k_{1}\)] (r2),
  };
       \end{feynman}
       \end{tikzpicture}
    \caption{The $Z$ mediated DM-electron scattering process.}
    \label{Z mediation}
\end{figure}

In addition,  with the small mass splitting between the lightest real and imaginary components of the new scalar fields, 
we can have the inelastic scatting process mediated by the $Z$ boson, as given in Fig.~\ref{Z mediation}. 
The corresponding matrix element can be written as
\begin{equation}
i M_{\text {free }}=\frac{\bar{u}(k)\left(\frac{1}{2} i\left(g_{2} \cos \theta_{W}-g_{1} \sin \theta_{W}\right)\left(\gamma^{\mu} \cdot \frac{1-\gamma_{5}}{2}\right)-i g_{1} \sin \theta_{W}\left(\gamma^{\mu} \cdot \frac{1+\gamma^{5}}{2}\right) u(p)\right) \cdot V_{SSZ}}{\left(k_{1}-p_{1}\right)^{2}-M_{Z}^{2}}~,
\end{equation}
\begin{equation}
V_{SSZ}=-\frac{1}{2} U_{12} U_{12}^{\prime }\left(g_{2} \cos \theta_{W}+g_{1} \sin \theta_{W}\right)\left(p_{1}-k_{1}\right)~,
\end{equation}
where $V_{SSZ}$, $U_{12}$, and $U_{12}^{\prime}$ are the vertex of $S_{1}S_{1}^{\prime}Z$,  the mixing matrix element of real, and imaginary parts of scalars, respectively. Also, $\theta_{W}$ is the Weinberg angle, 
and $g_{1}$ and $g_{2}$ are the gauge couplings of $U(1)_{Y}$ and $SU(2)_{L}$.

We find that the XENONnT data cannot give a constraint on the $Z$ mediated DM-electron scattering process
 since  $Z$ boson is heavy enough and the coupling of $S_{1} S_{1}^{\prime} Z$ is suppressed by mixing angles 
in both real ($U_{12}$) and imaginary parts ($U_{12}^{\prime}$) of scalars.

\section{Conclusion}\label{Section6}

We considered the general scalar potential with CP-violation and the inelastic dark matter
 in the Simple Extension of 
the Standard Model (SESM), which can explain the dark matter as well as
 new physics anomalies such as the B physics anomalies and muon anomalous magnetic moment, etc.
With the $a_{H}$, $a_{H}^{\prime}$ ${M_{S}^{\prime}}^{2}$, $\lambda_{SH}^{\prime}$,  and $\lambda_{DH}^{\prime\prime}$ terms, 
we obtained the mass splittings for the real and imaginary parts of scalar fields.
In particular, we can have the DM co-annihilation process mediated by $Z$ boson, which couples exclusively  
to the CP-even and CP-odd parts of scalar fields. For the CP conserving case, we presented 
the viable parameter space for the DM co-annihilation processes 
through both Higgs and $Z$ resonances, which can address the B physics anomalies, muon g-2, and DM relic density, as well as
evade the constraint from the XENON1T direct detection simultaneously. 
 For the CP-violating case, we discussed four scenarios for 
the inelastic DM-nucleon scatterings mediated by the Higgs and $Z$ bosons in details,
 and presented the benchmark points which satisfy the aforementioned constraints.
Finally, with the XENONnT results, we studied the inelastic scattering processes 
between the DM and electron mediated by Higgs and $Z$ bosons,
and found that the constraint on the $Z$ mediated process is weak, while the Higgs mediated process excludes 
the dark matter with mass around several MeV.

\appendix
\section{The Relevant Vertices}\label{A}
\begin{figure}[h]
        \centering
  \begin{tikzpicture}
       \begin{feynman}
       \vertex (m1);
       \vertex [above left=of m1](l1) {\(S_{i}\)};
       \vertex [below left=of m1](l2) {\(S_{j}\)};
       \vertex [ right=of m1](r2) {\(h\)};
  
       \diagram*{
  (l1) -- [scalar] (m1) -- [scalar] (r2),
  (l2) -- [scalar] (m1) ,
  };
       \end{feynman}
       \end{tikzpicture}
    \label{SiSjh}
\end{figure}
\begin{equation}
\begin{split}
    &\frac{i}{4}\left(4 v (2\lambda_{DH}^{\prime \prime} - \lambda_{DH} +\lambda_{DH}^{\prime})U_{i2}U_{j2} + 2\sqrt{2}\left(Re(a_{H}) - Re(a_{H}^{\prime})\right)U_{i2} U_{j1} \right. \\
    &\left.+4v\left(2Re(\lambda_{SH}^{\prime})-\lambda_{SH} \right)U_{i1}U_{j1} + 2\sqrt{2}\left(Re(a_{H}) - Re(a_{H}^{\prime})\right)U_{i1} U_{j2}        \right)~,
\label{A1}
\end{split}
\end{equation}
\newpage
\begin{figure}[h]
        \centering
  \begin{tikzpicture}
       \begin{feynman}
       \vertex (m1);
       \vertex [above left=of m1](l1) {\(S_{i}^{\prime}\)};
       \vertex [below left=of m1](l2) {\(S_{j}^{\prime}\)};
       \vertex [ right=of m1](r2) {\(h\)};
  
       \diagram*{
  (l1) -- [scalar] (m1) -- [scalar] (r2),
  (l2) -- [scalar] (m1) ,
  };
       \end{feynman}
       \end{tikzpicture}
    \label{SipSjph}
\end{figure}
\begin{equation}
\begin{split}
    &-\frac{i}{4}\left(4 v \left(2Re(\lambda_{SH}^{\prime}) + \lambda_{SH} \right)U_{i1}^{\prime}U_{j1}^{\prime} + 2\sqrt{2}\left(Re(a_{H}) + Re(a_{H}^{\prime})\right)U_{i1}^{\prime} U_{j2}^{\prime} \right. \\
    &\left.+4v\left(2\lambda_{DH}^{\prime \prime}-\lambda_{DH}^{\prime} + \lambda_{DH} \right)U_{i2}^{\prime}U_{j2}^{\prime} + 2\sqrt{2}\left(Re(a_{H}) + Re(a_{H}^{\prime})\right)U_{i2}^{\prime} U_{j1}^{\prime}        \right)~,
\label{A2}
\end{split}
\end{equation}

\begin{figure}[h]
        \centering
  \begin{tikzpicture}
       \begin{feynman}
       \vertex (m1);
       \vertex [above left=of m1](l1) {\(\mathscr{S}_{j}\)};
       \vertex [below left=of m1](l2) {\(\mathscr{S}_{k}\)};
       \vertex [ right=of m1](r2) {\(h\)};
  
       \diagram*{
  (l1) -- [scalar] (m1) -- [scalar] (r2),
  (l2) -- [scalar] (m1) ,
  };
       \end{feynman}
       \end{tikzpicture}
    \label{SiSjph}
\end{figure}
\begin{equation}
\begin{split}
    &\frac{i}{4}\left( -8 v U_{j3}U_{k1} Im(\lambda_{SH}^{\prime}) -2\sqrt{2}U_{j4}U_{k1}\left(Im(a_{H})+Im(a_{H}^{\prime}) \right) +2\sqrt{2}U_{j3}U_{k2}\left(Im(a_{H}^{\prime})-Im(a_{H})\right) \right. \\
    -&\left.4v U_{j3}U_{k3}\left(2Re(\lambda_{SH}^{\prime})+\lambda_{SH}\right) -2\sqrt{2}U_{j4}U_{k3}\left(Re(a_{H})+Re(a_{H}^{\prime})\right) + 4v U_{j2}U_{k2}\left(2\lambda_{DH}^{\prime\prime}-\lambda_{DH}+\lambda_{DH}^{\prime}\right) \right. \\
    +&\left.2\sqrt{2}U_{j2}U_{k1}\left(Re(a_{H})-Re(a_{H}^{\prime})\right)+2\sqrt{2}U_{j2}U_{k3}\left(Im(a_{H}^{\prime})-Im(a_{H})\right)-2\sqrt{2}U_{j3}U_{k4}\left(Re(a_{H})+Re(a_{H}^{\prime})\right)\right.\\
    +&\left.4vU_{j4}U_{k4}\left(\lambda_{DH}^{\prime}-2\lambda_{DH}^{\prime\prime}-\lambda_{DH}\right)-4vU_{j1}U_{k1}\left(\lambda_{SH}-2Re(\lambda_{SH}^{\prime})\right)+2\sqrt{2}U_{j1}U_{k2}\left(Re(a_{H})-Re(a_{H}^{\prime})\right)\right. \\
    -&\left.8vU_{j1}U_{k3}Im(\lambda_{SH}^{\prime})-2\sqrt{2}U_{j1}U_{k4}\left(Im(a_{H})+Im(a_{H}^{\prime})\right)\right)
\label{A3}
\end{split}
\end{equation}

\begin{figure}[h]
        \centering
  \begin{tikzpicture}
       \begin{feynman}
       \vertex (m1);
       \vertex [above left=of m1](l1) {\(S_{i}\)};
       \vertex [below left=of m1](l2) {\(S_{j}^{\prime}\)};
       \vertex [ right=of m1](r2) {\(Z\)};
  
       \diagram*{
  (l1) -- [scalar] (m1) -- [boson] (r2),
  (l2) -- [scalar] (m1) ,
  };
       \end{feynman}
       \end{tikzpicture}
    \label{SiSjph}
\end{figure}
\begin{equation}
    -\frac{1}{2}\left( g_{1} \text{sin}\theta_{W} +g_{2} \text{cos} \theta_{W} \right)U_{i2}U_{j2}^{\prime}\left(-p_{\mu}^{S_{j}^{\prime}}+p_{\mu}^{S_{i}}\right)~.
\label{A4}
\end{equation}



\acknowledgments

We are indebted to Lorenzo Calibbi, Jibo He, Junle Pei, and Bin Zhu for the helpful discussions. 
This research is supported in part by the National
Key Research and Development Program of China Grant No. 2020YFC2201504, by the
Projects No. 11875062, No. 11947302, No. 12047503, and No. 12275333 supported by
the National Natural Science Foundation of China, by the Key Research Program of the
Chinese Academy of Sciences, Grant No. XDPB15, by the Scientific Instrument Developing
Project of the Chinese Academy of Sciences, Grant No. YJKYYQ20190049, and by the
International Partnership Program of Chinese Academy of Sciences for Grand Challenges,
Grant No. 112311KYSB20210012.



\end{document}